\documentstyle[aps,twocolumn,graphics]{revtex}
\newcommand{\andy}[1]{}
\begin{document}
\title{Designing optimum CP maps for quantum teleportation}
\author{J. \v{R}eh\'{a}\v{c}ek$^1$\thanks{e-mail: 
rehacek@alpha.inf.upol.cz}, 
Z. Hradil$^1$, J. Fiur\'{a}\v{s}ek$^1$, and \v{C}. Brukner$^2$}
\address{$^1$ Department of Optics, Palacky University, 
17. listopadu 50, 772~00 Olomouc, Czech Republic\\
$^2$ Institute for Experimentalphysics, University of Vienna,
Boltzmanngasse 5, A-1090, Vienna, Austria}
\maketitle

\begin{abstract}
We study general teleportation scheme with an arbitrary 
state of the pair of particles ($2$ and $3$)
shared by Alice and Bob, and arbitrary measurements on the
input particle 1 and one of the members (2) of the pair on 
Alice's  side. We find an efficient iterative algorithm for 
identifying optimum local operations on Bob's side.
In particular we find that simple unitary transformations
on his side are not always optimal even if particles $2$ and 
$3$ are perfectly entangled. We describe the most interesting 
protocols in the language of extremal completely-positive maps.
\end{abstract}

\vspace{0.5cm}

Of many potential applications of quantum information processing,
the quantum teleportation is probably the most appealing example.
Several experiments \cite{experiment} have been done 
since the first proposal in \cite{proposal},
confirming thus that teleportation -- a popular
subject of sci-fi literature -- is indeed
feasible; at least if one deals with simple
quantum objects.

It is well known that ideal teleportation requires
a source of maximally entangled particles and 
a very delicate measurement on the sender's side.
This is not always easy to do with realistic experimental
devices. For example, it has been shown 
in \cite{bell-impossible} that a never-failing Bell-state
measurement is impossible with linear elements and detectors only.
More severe limitations might arise
if the teleported object gets more complicated. 

However, even with restricted resources there is still the
possibility to optimize the teleporting scheme.
A natural question is: What is the optimum local
operation on particle $3$ (output of teleportation)
for the given resources of the shared entangled pairs of particles 
($2$ and $3$), and of the measurements performed jointly on particle $2$
and the input particle $1$?
Several attempts have been done in this direction.
In \cite{realistic} the authors analyzed a teleportation
with realistic linear elements. They found that 
local unitary transformations on the side of receiver
are not always optimum; sometimes more general operations
can improve the performance of the realistic protocol.
Optimization of protocols with arbitrary shared entangled states 
has been pursued in \cite{3H} and \cite{banaszek}
but the optimization was done over local unitary 
transformations only.

The main goal of this paper is to consider the most general case. 
For the given resources of the sender 
(type of the measurement on particles $1$ and $2$) and 
resources on the quantum channel shared by two parties
(the state of particles $2$ and $3$) we will optimize
the teleportation protocol by finding the optimum 
local operation that should be applied by the receiver
on particle $3$.
In contrast to the naive picture suggesting that
every interaction of a quantum system with 
environment leads to a ``loss of information'',
we here find that such an interaction on receiver's
side might enhance the fidelity of teleportation even 
if the pair of particles constituting the quantum channel
are perfectly entangled.

The paper has two parts. In the first one we will
derive an iterative algorithm that solves our problem.
In the second part we will try to describe the most
interesting protocols involving two-level systems
in the language of extremal completely-positive maps.

Let us consider the teleportation of an 
unknown state $\bar{\rho}_1$
of particle $1$ between two parties called Alice and Bob. 
Let us assume that before the teleportation starts they
share two particles in an arbitrary state $\tau_{23}$.
Alice then performs a measurement on particles $1$ and $2$
and sends the outcome by a classical channel to Bob.
Based on the classical communication he receives he performs a 
local transformation on particle $3$. The optimum 
local transformation is such that the final state of 
particle $3$ gets as close to the input state $\bar{\rho}_1$ as 
possible on the average.

The measurement performed by Alice will be described by a
positive operator-valued measure (POVM) $\{\Pi^j_{12}\}$, 
$\sum_j \Pi^j_{12}=1$. 
Its elements generate probabilities of all possible 
outcomes of Alice's measurement.
Consider the situation where one such 
particular outcome, say $j=a$ has been registered
with probability $p_a$.
The state of the third particle conditioned on this result
reads
\andy{state-cond}
\begin{equation}\label{state-cond}
\rho^{a}_{3}=\frac{1}{p_{a}}{\rm Tr}_{12}\left\{ 
\bar{\rho_1} \tau_{23} \Pi_{12}^{a}\right\}=
\frac{1}{p_a}{\rm Tr}_1\left\{\bar{\rho}_1 O^a_{13}\right\},
\end{equation}
where
\andy{O13}
\begin{equation}\label{O13}
O^a_{13}={\rm Tr}_{2}\{\tau_{23}\Pi^a_{12}\}.
\end{equation}
Now Bob applies a local ($a$-dependent) 
transformation on this state. The most general
transformation is a trace preserving completely positive (CP) map
of the form \cite{kraus}
\andy{cp}
\begin{equation}\label{cp}
\Phi^a(\rho_3^a)=\sum_{k}A^a_{k3}\rho_3^a A_{k3}^{a\dag},\quad 
\sum_{k}A^{a\dag}_{k3}A^a_{k3}=1.
\end{equation}
The state describing the ensemble of teleported particles 
is obtained by averaging (\ref{cp}) over all possible outcomes,
\andy{final}
\begin{equation}\label{final}
\rho_3=\sum_a \sum_k p_a A_{k3}^a \rho_3^a A_{k3}^{a \dag}.
\end{equation}
The optimality of the given set of local transformations will be
judged by the fidelity of the corresponding teleported state.
To simplify further considerations let us assume that the input 
state is pure \cite{note-pure}. In that case the fidelity can be defined 
as follows: $F={\rm Tr}\{\bar{\rho}_3 \rho_3\}$, where
$\bar{\rho}_3$ is the input state in the Hilbert space of 
particle $3$. 
Using Eqs.~(\ref{state-cond}) and (\ref{final}) the fidelity 
becomes
\andy{fidel}
\begin{equation}\label{fidel}
\langle F\rangle=\sum_a {\rm Tr}_{13}\left\{ 
\langle \bar{\rho}_3 \bar{\rho}_1\rangle 
\sum_k A_{k3}^a O^a_{13} A^{a\dag}_{k3}\right\},
\end{equation}
where $\langle \dots \rangle$ mean averaging over
a priori distribution of input states. 
The average fidelity $\langle F\rangle$ will be our global 
measure of quality of the given teleportation protocol that
is to be maximized over the set of all possible local operations
applied to particle $3$.

Although the following optimization can be carried out for arbitrary 
dimensional Hilbert spaces we will illustrate the idea on the
simple example of spin-$1/2$ particles. The generalization to
more dimensions is straightforward.

We start by decomposing the input density matrix in some basis of 
Hermitian generators. In the case of spin $1/2$ the convenient choice
is the basis of Pauli spin matrices. After substituting the decomposition 
in Eq.~(\ref{fidel}) and simple integration over the whole surface of
Poincare sphere (we assume that all possible 
input states are equally-probable
states, hence the isotropic a priori distribution), we get \andy{f2}
\begin{equation}\label{f2}
\langle F\rangle =\frac{1}{2}+\frac{1}{12}\sum_a\sum_k
{\rm Tr}\left\{ \vec{\sigma }
A^{a}_{k}\vec{O}^{a} A^{a\dag }_{k}\right\},
\end{equation}
where
\andy{O}
\begin{equation}\label{O}
\vec{O}^a={\rm Tr}_{1}\{\vec{\sigma}_1 O_{13}^a\},
\end{equation}
where $\vec{\sigma}=(\sigma_x,\sigma_y,\sigma_z)$ and 
where we have dropped now unnecessary subscript of particle $3$.

Since CP maps corresponding to different registrations \( a \) are 
independent, each term on the right-hand side of (\ref{f2}) 
can be maximized independently.
Omitting therefore notation $a$ and taking the 
constraint \( \sum _{k}A^{\dag }_{k}A_{k} =1\) into account, 
the expression to be maximized is
\andy{tobemax}
\begin{equation}\label{tobemax}
\label{cp-func}
\sum_k {\rm Tr}\left\{ \vec{\sigma} A_{k}\vec{O} A_{k}^{\dag }-
A_{k}\Lambda A_{k}^{\dag }\right\} =max.
\end{equation}
Variation of this expression with respect to $ A_{k}^{\dag } $ gives
the extremal equation in the form
\andy{cp-extrem}
\begin{equation}
\label{cp-extrem}
\sigma _{i}A_{k}O_{i}=A_{k}\Lambda,
\end{equation}
where $\Lambda$ is the (hermitian) Lagrange operator.
It can be determined from Eq.~(\ref{cp-extrem}) as follows: 
\andy{cp-lagrang}
\begin{equation}
\label{cp-lagrang}
\Lambda =(\vec{X}\cdot \vec{O}+\vec{O}\cdot \vec{X})/2,
\end{equation}
where we have introduced hermitian operators
\andy{X}
\begin{equation}\label{X} 
\vec{X}=\sum _{k}A^{\dag }_{k}\vec{\sigma }A_{k}
\end{equation}
which provide another representation of the CP map $\{A_k\}$.
Eq.~(\ref{cp-extrem}) can be brought to the form
suitable to iterations. Multiplying it by \( A_{k}^{\dag }\sigma _{j} \) 
from the left and summing over \( k \) we obtain
\andy{cp-vec}
\begin{equation}
\label{cp-vec}
\vec{X}\Lambda =\vec{O}-i|\vec{X}\times \vec{O}|,
\end{equation}
A formula suitable
to iterative solving is obtained by adding 
\( \vec{0}=\vec{X}-\vec{X} \) to the left-hand
side of Eq.~(\ref{cp-vec}) and rearranging,
\andy{iter}
\begin{equation}\label{iter}
\vec{X}=\vec{X}+\vec{O}-\left(i|\vec{X}\times \vec{O}|+\vec{X}\Lambda +
{\rm H.c.}\right). 
\end{equation}
The iterative algorithm for finding optimum CP maps 
based on Eqs.~(\ref{cp-lagrang}) and 
(\ref{iter}) is the main formal result of the 
present article. Starting from some ``unbiased'' CP map, for example
$\vec{X}=\vec{0}$ (means that particle $3$ is always 
brought to the maximally mixed state), the equations can be successively 
iterated until the stationary point is attained. 
In this way we get the operators $\vec{X}$ corresponding to the optimum 
local transformation of particle $3$. 

Notice that the average fidelity of teleportation bears very simple form
when expressed in terms of \( \vec{X} \),
\andy{f3}
\begin{equation}\label{f3}
\langle F\rangle =\frac{1}{2}+\frac{1}{12}\sum _{a}{\rm Tr}
\left\{\vec{X}^{a}\cdot \vec{O}^{a}\right\}.
\end{equation}
Notice also that $\langle F\rangle$ is a linear functional of $\vec{X}$.
This means that all its maxima lie on the boundary of the
set of physically allowed operators $\vec{X}$ that is 
determined by the constraint of complete positiveness of the 
corresponding local transformations. So there is a clear connection 
between optimum teleportation protocols and extremal CP maps.
The topology of CP maps is a well studied field related to many
problems in quantum information processing. We
will use some of the recently derived results on CP maps 
for the discussion of some cases of special interest
in the present problem.

But before we come to this point let us first demonstrate
the usefulness of our iterative optimizing algorithm on some 
interesting examples involving spin $1/2$ systems.
It is well known that ideal teleportation protocols
($\langle F\rangle=1$) utilize a source of maximally entangled 
pairs of particles and a Bell state analysis on Alice's side.
The latter is definitely more difficult to realize in a laboratory.
Let us therefore analyze the realistic situation with an (almost) perfect  
source of shared particles but with an imperfect measurement.
The imperfect measurement will be drawn from this one-parametric
family of POVMs,
\andy{POVMs}
\begin{eqnarray} \label{POVMs}
\Pi_{12}^a&=&\frac{\sin^2\theta}{2}|--\rangle\langle --|
+\frac{1}{2}|\phi^a\rangle\langle \phi^a|,\nonumber \\
&&|\phi^a\rangle=\cos\theta|+-\rangle-|-+\rangle,\nonumber\\
\Pi_{12}^b&=&\frac{\sin^2\theta}{2}|++\rangle\langle ++|
+\frac{1}{2}|\phi^b\rangle\langle \phi^b|,\nonumber \\
&&|\phi^b\rangle=\cos\theta|-+\rangle+|+-\rangle,\\
\Pi_{12}^c&=&\frac{\sin^2\theta}{2}|+-\rangle\langle +-|
+\frac{1}{2}|\phi^c\rangle\langle \phi^c|,\nonumber \\
&&|\phi^c\rangle=\cos\theta|--\rangle+|++\rangle,\nonumber \\
\Pi_{12}^d&=&\frac{\sin^2\theta}{2}|-+\rangle\langle -+|
+\frac{1}{2}|\phi^d\rangle\langle \phi^d|,\nonumber \\
&&|\phi^d\rangle=\cos\theta|++\rangle-|--\rangle.\nonumber
\end{eqnarray}
Here $|+\rangle$ and $|-\rangle$ are two orthogonal states,
for example states spin up and spin down in the $z$ direction
and $\theta\in [0,\pi/2]$.
One boundary point $\theta=0$ corresponds
to the perfect Bell measurement. Alice gets no information
about the incoming state but the teleportation 
with fidelity one is possible.
The other boundary point $\theta=\pi/2$ corresponds to a 
projective measurement on the first particle and no measurement 
on the second one. In this case 
$\Pi^a=\Pi^d$ and $\Pi^b=\Pi^c$, and
we have only two distinct outcomes $\Pi^a+\Pi^d=
|-\rangle\langle-|_1\otimes \hat{1}_2$ and $\Pi^b+\Pi^c=
|+\rangle\langle+|_1 \otimes\hat{1}_2$. 
Alice gets maximum amount of information about the input state,
but the ``teleported'' state bears no relation to the input state
-- the quantum resources are wasted in this case.

For intermediate values of $\theta$ less information  is extracted
about the input state and  an imperfect quantum
teleportation is possible. 
As we have mentioned above we will take particles $2$ and $3$
in a maximally entangled state, for instance let them be
in the singlet state
\andy{max-ent}
\begin{equation}\label{max-ent}
\tau_{23}=\frac{1}{4}(1-\sigma_{1x}\sigma_{2x}-
\sigma_{1y}\sigma_{2y}-\sigma_{1z}\sigma_{2z}).
\end{equation}
Accordingly, the operators $\vec{O}^a$ have the form,
\andy{Hor-O}
\begin{equation}\label{Hor-O}
\vec{O^a}=4 R\vec{\sigma}+4\vec{r},
\end{equation}
where 
\andy{Rr}
\begin{equation}\label{Rr}
R=\left(
\begin{array}{ccc}
\cos\theta&0&0\\
0&\cos\theta&0\\
0&0&\cos^2\theta
\end{array}\right),\quad
\vec{r}=\left(
\begin{array}{c} 0\\0\\-\sin^2\theta\end{array}\right).
\end{equation}
The operators $\vec{O}$ generated by the remaining
three POVM elements $\Pi^b$, $\Pi^c$ and $\Pi^d$ 
differ from (\ref{Rr}) only in signs of its elements and hence
their contribution to the average fidelity (\ref{f3})
is the same. 
 
\begin{figure}
\centerline{
\scalebox{0.45}{\includegraphics{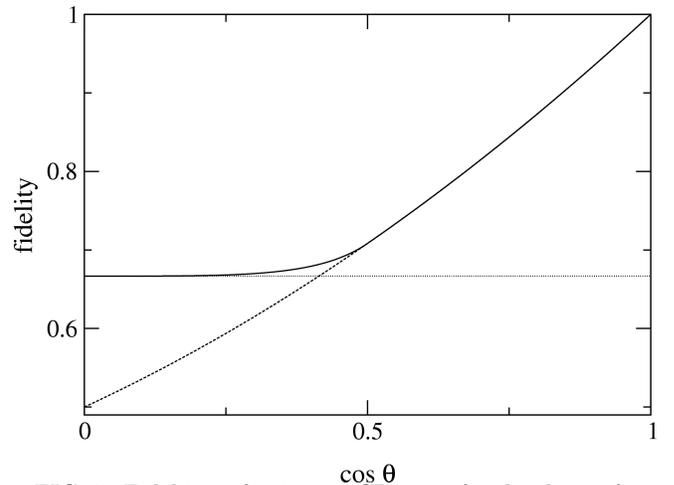}}
}
\caption{Fidelities of optimum CP maps for the chosen 
family of Alice's measurements (\ref{POVMs}). Solid line
shows the performances of optimum CP maps; dashed line
shows the performances of optimum local unitary operations.
Dotted horizontal line shows the boundary between classical and
quantum teleportation protocols.}
\label{fig-fidel}
\end{figure}

Fidelities (\ref{f3}) of the optimum Bob's local transformations
that were found by our iterative algorithm (\ref{cp-lagrang})
and (\ref{iter}) for the above Alice's measurements
are shown in Fig.~\ref{fig-fidel} (solid line).
As could have been anticipated, the fidelity continuously changes
from the classical limit $\langle F\rangle=2/3$ 
\cite{classical} to the maximum 
value of $\langle F\rangle=1$. It is interesting to note
that the optimum CP maps for $\cos\theta\ge 1/2$ are actually  
local unitary operations. However for $\cos\theta<1/2$
unitary operations are not optimum, see 
Fig.~\ref{fig-fidel} (dashed line).
This becomes trivial statement if $\cos\theta=0$.
In that case Alice performs no measurement on particle $2$
and therefore the state of particle $3$ remains a complete 
mixture for any outcome $a$ she obtains. So
the fidelity of ``teleported'' state is 
$1/2$ if one is allowed to perform unitary transformations
only. In contrast, if one adopts more general transformations,
Bob can construct the input state with fidelity $2/3$
on the basis of the outcomes of Alice's measurement
(which is here optimal state estimation) thus attaining 
the classical limit.

The most interesting situations correspond to Alice's 
measurements that allow non-classical teleportation only if
followed by a {\em non unitary} operation on Bob's particle.
In our case this happens for $0<\cos\theta\le\sqrt{2}-1$ 
(see later) and the optimum CP map turns out to be
a kind of decoherence process \cite{horod}.

Let us emphasize again that the above simple example of teleporting
spin $1/2$ system has been chosen for the sake of simplicity only.
If needed, our iterative algorithm could be straightforwardly 
generalized to larger Hilbert spaces. The generalization consists
of replacing Pauli spin matrices by the appropriate basis of 
hermitian operators in that space and replacing the 
integration over the surface  of the Poincare sphere by the 
integration over the surface of generalized  $N$-dimensional
sphere. 

Nevertheless, the relative simplicity of the set of CP maps
operating on a $2$-dimensional space allows one to get 
further insight into the optimum 
teleportation protocols and obtain analytical results.
We will therefore stick to the teleportation of 
spin $1/2$ particles in the following.

First let us express the local operation acting on the Bob's 
particle using operators $\vec{X}$,
\andy{lo}
\begin{equation}\label{lo}
\Phi(1/2+\vec{w}\cdot\vec{\sigma}/2 ) =1/2+(\vec{t}+T\vec{w})
\cdot\vec{\sigma},
\end{equation}
where $\vec{t}$ and $T$ are defined as follows:
$\vec{X}=T\vec{\sigma}+\vec{t}$, and $\vec{w}$ is the Bloch vector
defining the state of the Bob's particle before he applies 
the operation $\Phi$.
To each local operation on the Bob's particle there correspond 
a simple transformation of the Poincare sphere:
The unit sphere is mapped onto an ellipsoid, the lengths of its axes
being the eigenvalues of $T$, which is translated by $\vec{t}$ 
from the origin. Of course the ellipsoid has to lie within
the unit Poincare sphere (positivity). 
However not all such ellipsoids define {\em completely} positive maps
which are the most general maps in quantum mechanics.
Recently it has been shown \cite{ruskai} how to parameterize the set 
of {\em extremal} CP maps comprising the boundary of the convex set of all 
CP maps. This set contains all optimum
Bob's local transformation. Matrix $T$ can be always brought to the 
diagonal form by a local unitary transformation. When in diagonal form,
extremal CP maps  can be parametrized by two angles $u$ and $v$,
\andy{uv}
\begin{equation}\label{uv}
T=\left(
\begin{array}{ccc}
\cos u& 0& 0\\
0 &\cos v &0\\
0 & 0& \cos u \cos v
\end{array}\right),\quad
\vec{t}=\left(
\begin{array}{c}
0 \\0 \\ \sin u \sin v
\end{array}\right)
\end{equation}  
with $u\in[ 0,2\pi)$, $v\in[ 0,\pi)$.  
Now let us show how the above example can be elegantly solved 
using this trigonometric parameterization of extremal CP maps.
Due to the form of matrix $R$ Eq.~(\ref{Rr}) it can be shown 
that the optimum CP map is degenerated, $u=2\pi-v$, so there 
is only one free parameter left (see Appendix). 
  
Substituting the map (\ref{uv}) with $u=2\pi-v$ together with 
the POVM element
(\ref{Rr}) into Eq.~(\ref{f3}) and maximizing the fidelity
with respect to $v$ one easily finds the analytical
expression for the optimum fidelity:
\andy{opt-f}
\begin{equation}\label{opt-f}
\langle F_{\rm opt}\rangle=\left\{
\begin{array}{ll}
\frac{\cos^4\theta-4\cos^2\theta+2}{3-6\cos^2\theta}, &
\cos\theta\in [ 0, \frac{1}{2})\\[1.5mm]
(\cos^2\theta+2\cos\theta+3)/6,  & \cos\theta\in [ \frac{1}{2}, 1]
\end{array} 
\right.
\end{equation} 
In contrast to this, the fidelity of the optimum local unitary operation 
is given by the second of the two expressions for all angles $\theta$ and
thus coincides with the optimum value if $\cos\theta\ge 1/2$.
We note that the optimum unitary operation is the identity map 
$u=v=0$ (Bob leaves particle $3$ alone) for all $\theta$.
The analytical solution confirms the results obtained 
numerically with the help of our 
iterative algorithm shown in Fig.~\ref{fig-fidel}.
The most interesting Alice's measurement (\ref{POVMs}) is that
for which optimum unitary operation just gives 
the classical limit $\langle F\rangle=2/3\approx 0.667$, but enables 
non-classical teleportation using a non-unitary CP map.
This happens for $\cos\theta=\sqrt{2}-1$, optimum fidelity being
$\langle F_{\rm opt}\rangle=(3+8\sqrt{2})/21\approx 0.6816.$

The enhancement of quantum teleportation by non-unitary process 
has recently been discussed by Badzi\c ag and Horodeckis 
family \cite{horod}  in a slightly different context.
They considered teleportation protocols with perfect Bell
analyzer but imperfect preparation of the shared pair of particles and
found out that teleportation protocols can sometimes be enhanced by
the interaction of the teleportation device with environment (damping).
The most pronounced example yielded improvement corresponding to
our $\cos\theta=\sqrt{2}-1$ case. In fact the choice of 
our POVMs (\ref{POVMs}) which are Bell states being subject to a kind
of decohering process has been inspired by their result. 
Now we can use the language of extremal CP maps to explain
why the result of Badzi\c ag {\em et al.} is so exceptional.
A general teleportation protocol is, in fact, a CP map composed
of two different maps: The first one $\Psi$ being the Alice's 
transformation of the input state to the output state sent to Bob; 
the second being the local operation $\Phi$ on the output state 
applied by Bob. The teleportation protocol $\Omega$
becomes perfect if the two maps  make up the identity map:
\andy{cp-unit}
\begin{equation}\label{cp-unit}  
\Omega(\bar{\rho}_1)\equiv \sum_a p_a\Phi^a[\Psi^a(\bar{\rho}_1)]=
\bar{\rho}_3,\quad \forall \bar{\rho}_1.
\end{equation}
Obviously, the most interesting protocols are protocols where
both parts $\Psi$ and $\Phi$ are {\em extremal} CP maps,
because they contain optimal maps, see discussion after
Eq.~(\ref{f3}), and because all remaining protocols 
are just convex combinations of such ``extreme'' protocols. 
Among the protocols consisting of two extremal maps one can find
the standard teleportation, both maps here being  
unitary operations, but this is also the case of our  
example (\ref{POVMs})! 
This is immediately seen using another equivalent 
representation of our $\Psi$ \cite{sacchi}:
\andy{another}
\begin{equation}\label{another}
\Psi(\bar{\rho}_1)={\rm Tr}_1\left\{\bar{\rho}^T_1 
\chi_{13}\right\},
\end{equation}
where $\chi_{13}=O^{\tilde{T}}_{13}$, and $\tilde{T}$ is partial 
transposition with respect to system $1$.  
For POVMs from our example (\ref{POVMs}) and shared singlets 
(\ref{max-ent}) we have that the operator $\chi$ is at most 
rank $2$ operator and hence the map $\Psi$ it generates is 
extremal. Also, the associated operators $\vec{O}$, 
entering the fidelity expression (\ref{f3}) in a similar 
way as $\vec{X}$ do, posses the  form of (\ref{uv}). 

The trigonometric parameterization of optimum CP maps hints
on a possible generalization of Horodecki's considerations.
One could think of a situation where the first CP map $\Psi$
would not be degenerated with respect 
to angles $u$ and $v$ [unlike in (\ref{Rr})]. Such non-degenerated
Alice's POVMs would not however lead to substantially new physics
since also in this case the optimum Bob's operation would be 
realizable by a unitary operation on an enlarged space of the 
third qubit followed by tracing out the auxiliary 
degrees of freedom -- thus representing a damping channel 
\cite{ruskai}.

The trigonometric parameterization (\ref{uv}) also provides
a nontrivial lower bound on the fidelity of teleportation.
Let us assume that Bob, for the lack of invention, 
decides just to repeat the first CP map which results from Alice's
measurement in the sense $\vec{X}=\vec{O}$.
In non-degenerated case this simple protocol yields fidelity
\andy{f-square}
\begin{equation}\label{f-square}
\langle F_{\rm repet}\rangle=\frac{\cos^2 u\, \sin^2 v+2}{3}\le
\langle F_{\rm opt}\rangle,
\end{equation}
and angles $u$ and $v$ define both $\vec{O}$ and $\vec{X}$.
This results shows another interesting, but not surprising, 
feature of extremal CP maps:
if used by both parties they cover the whole range of 
non-classical fidelities.

In conclusion, we have derived an iterative algorithm for 
finding optimum CP maps for quantum teleportation.
It provides a convenient way of optimizing 
teleportation protocols on Hilbert spaces of arbitrary
dimensions. We have applied our procedure to protocols with
imperfect Bell analysis and have identified situations where
a unitary transformation on the third particle is not optimal
and should be replaced by a more general 
completely-positive map.

{\bf Acknowledgements}: This work was supported by
grant no. LN00A015 of the Czech Ministry of Education and by
the program ``Quantum Measurement Theory and Quantum Information''
of ESI in Vienna. \v{C}. B. acknowledges the support by
QIPC program of the European Union.

\appendix

\section*{Degenerated extremal CP maps}

Using Eqs.~(\ref{Rr}) and (\ref{uv}) the relevant part of the
contribution to fidelity comming from Alice's registration $\Pi^a$
and Bob's CP map defined by angles $u$ and $v$ reads
\begin{eqnarray}
F'=Tr(RT)+\vec{r}\cdot\vec{t}=\cos u'\cos\theta +\cos v\cos\theta
\nonumber \\
+\cos u'\cos v\cos^2\theta+\sin u'\sin v \sin^2\theta,
\end{eqnarray} 
where $u'=2\pi-u$ and all terms are made positive by 
restricting the ranges of $u'$ and $v$ to $[0,\pi/2]$.

Now $F'$ is bounded from above by
\begin{eqnarray}
F'\le F''&=& 2\frac{\cos u'+\cos v}{2}\cos\theta+
\left(\frac{\cos u'+\cos v}{2}\right)^2\nonumber \\
&&\times\cos^2\theta
+\left(\frac{\sin u'+\sin v}{2}\right)^2\sin^2\theta
\end{eqnarray}
(arithmetic vs. geometrical mean), and this in turn is smaller 
than 
\begin{eqnarray}
F'\le F''\le F'''&=&2\cos(\frac{u'+v}{2})\cos\theta+
\cos^2(\frac{u'+v}{2})\nonumber\\
&&\times
\cos^2\theta+\sin^2(\frac{u'+v}{2})\sin^2\theta
\end{eqnarray}
(convexity). $F'''$ can be attained by applying 
degenerated CP map defined by angles $\bar{u}$, $\bar{v}$:
$\bar{u}=2\pi-\bar{v}=(u'+v)/2$.
Hence for Alice's POVMs (\ref{POVMs}) and the singlet state
shared by Alice and Bob the optimum Bob's CP map is contained
in the set of one-parametric CP maps.


\begin{thebibliography}{99}
\bibitem{note-pure} No generality is lost by this assumption
since mixed states are just convex combinations of 
pure states.
\bibitem{classical} Alice performes optimal state estimation
and sends this to Bob who then reconstructs the state on the basis
of this information.
\end{thebibliography}
\end{document}